\documentclass[a4paper,11pt]{article}
\usepackage{pos}

\title{Measurements of Nuclear Modification Factors of $B^{0}_{s}$ and $B^+$ Mesons in PbPb Collisions with the CMS Experiment}

\manuallySeparateAuthors
\author*[a]{Zhaozhong Shi}
\author{on behave of the CMS collaboration}

\affiliation[a]{Laboratory for Nuclear Science, Massachusetts Institute of Technology,\\
77 Massachusetts Avenue, Cambridge, Massachusetts, USA}

\emailAdd{zzshi@mit.edu}

\abstract{Beauty quarks are considered as one of the best probes of the strongly interacting medium created in relativistic heavy-ion collisions because they are predominantly produced via initial hard scatterings. Measurements of B meson production provide information about the diffusion of beauty quarks and the flavor dependence of in-medium energy loss. In these studies, clarifying the hadronization mechanism is crucial for understanding the transport properties of beauty quarks. Measurements of $B^{0}_{s}$ production can shed light on the mechanisms of beauty recombination in the medium and provide information about strangeness enhancement in the quark-gluon plasma. In this talk, we will present a new measurement of the ratio of $B^{0}_{s}$ to $B^+$ mesons in PbPb collisions at 5.02 TeV with the CMS detector, using data recorded in 2018.}

\FullConference{%
  HardProbes2020\\
  1-6 June 2020\\
  Austin, Texas}

\begin{document}

\maketitle

\section{Introduction}

Heavy quarks, such as charm quarks and beauty quarks, are considered as excellent probes to study the properties of the strongly interacting quark-gluon plasma (QGP) created in high-energy nuclei collisions. Due to their large masses, which are in the order of a few GeV/$c^2$, they are predominantly created in the initial hard scattering processes in relativistic heavy-ion collisions and can be calculated in perturbative QCD (pQCD). Because they traverse through the QGP medium before decaying, they can record the QGP evolution \cite{HFPhysics}. In addition, heavy quarks have weaker coupling strengths to the medium compared to light quarks \cite{HQCouple}. Therefore, they have long thermal relaxation time and are in general not completely thermalized in the QGP medium. 

Because Heavy quarks interact with the constituents in the QGP medium and lose significant amount of energy, they can be used to probe the QGP and to study medium properties from the interaction. In the pQCD picture, the energy loss of heavy quarks is described by two mechanisms: collisional energy loss and radiative energy loss. In the collisional process, heavy quarks transfers their momenta to the medium via $Qq \rightarrow Qq$ and $Qg \rightarrow Qg$  scattering (Q: heavy quark, q: light quark, g: gluon) \cite{Collisional}. It is the dominating factor of heavy quarks energy loss in the non-relativistic limit. In the radiative process, heavy quarks radiate gluons. The gluon radiation spectrum is given by $dP = \frac{\alpha_s C_F}{\pi} \frac{d\omega}{\omega} \frac{k_\perp^2 dk_\perp^2}{(k_\perp^2 + \omega^2 \theta_0^2)^2}$ where $\theta_0 = \frac{m}{E}$ \cite{DeadCone}. Hence, the flavor dependence of parton energy loss is expected to follow: $\Delta E_g > \Delta E_q > \Delta E_Q$.

The hadronization of parton is in general non-perturbative. Several phenomenological models such as the Statistical Hadronization Model \cite{SHM}, Lund String Model \cite{LSM}, and Quark Coalescence Model \cite{QCM} are applied to describe heavy quark hadronization in different conditions. Because the temperature of the thermally and chemically equilibrated QGP is higher than 150 MeV and greater than the mass of strange quarks, strange quarks can be thermally produced in the QGP medium via the process $gg \rightarrow s \bar s$. Hence, strangeness enhancement is expected to be observed in charm and beauty hadrons via the recombination hadronization mechanism \cite{Recombination}. 

We report the experimental observables of beauty mesons measured by the CMS experiment \cite{CMSDetector} to study the properties of the QGP. We will show the $B^+$ and $B_{s}^0$ nuclear modification factor ($R_{AA}$) vs $p_T$ measurements from the 2015 pp and PbPb datasets to study the beauty energy loss. $B_{s}^0$/$B^+$ as a function of transverse momentum and event centrality measurements from the 2018 PbPb dataset will also be presented to investigate the beauty hadronization mechanism in the QGP. 

\section{2015 B-meson Nuclear Modification Factor Results}

\begin{figure}[hbtp]
\begin{center}
\includegraphics[width=0.30\textwidth]{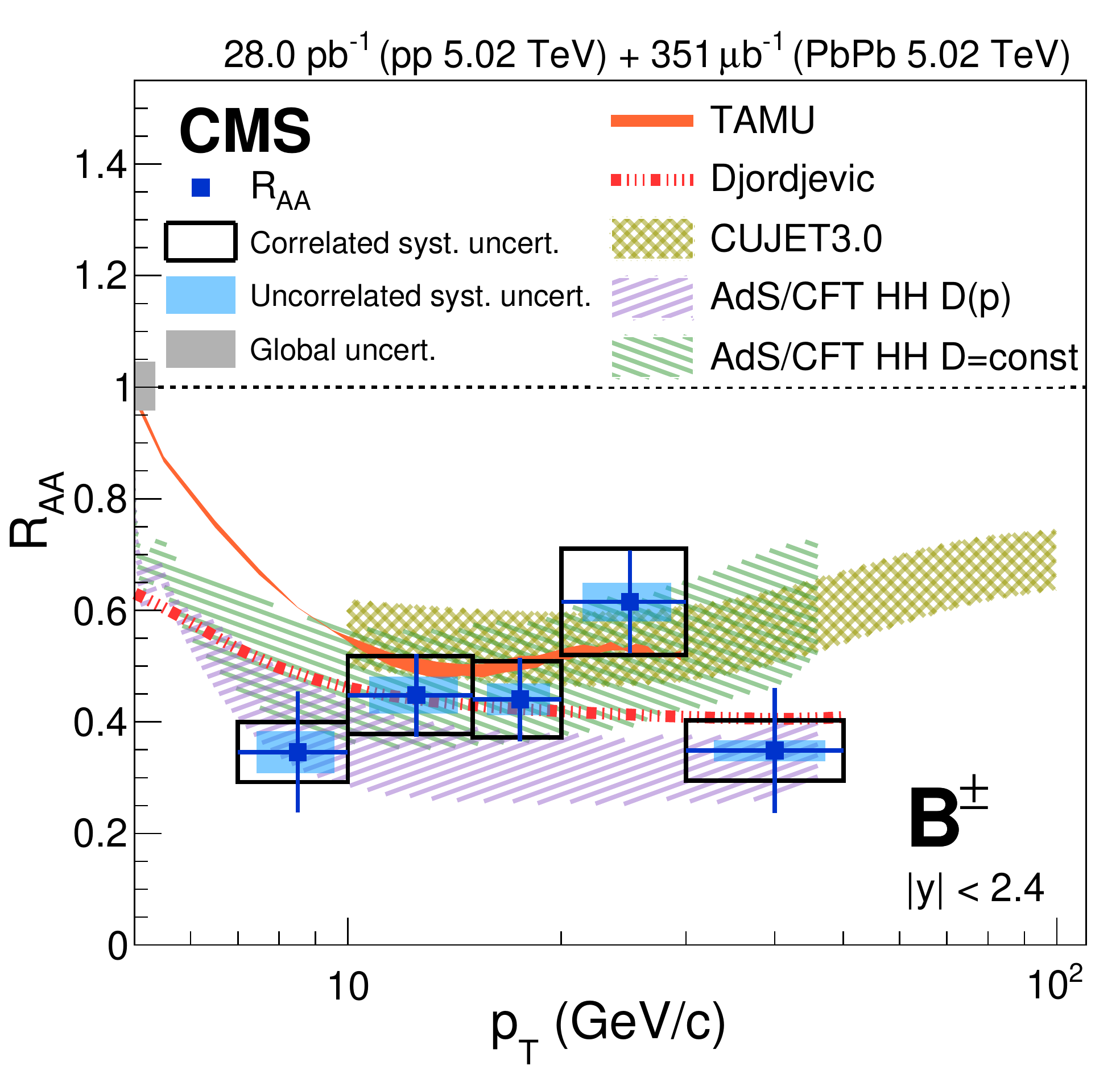}
\includegraphics[width=0.30\textwidth]{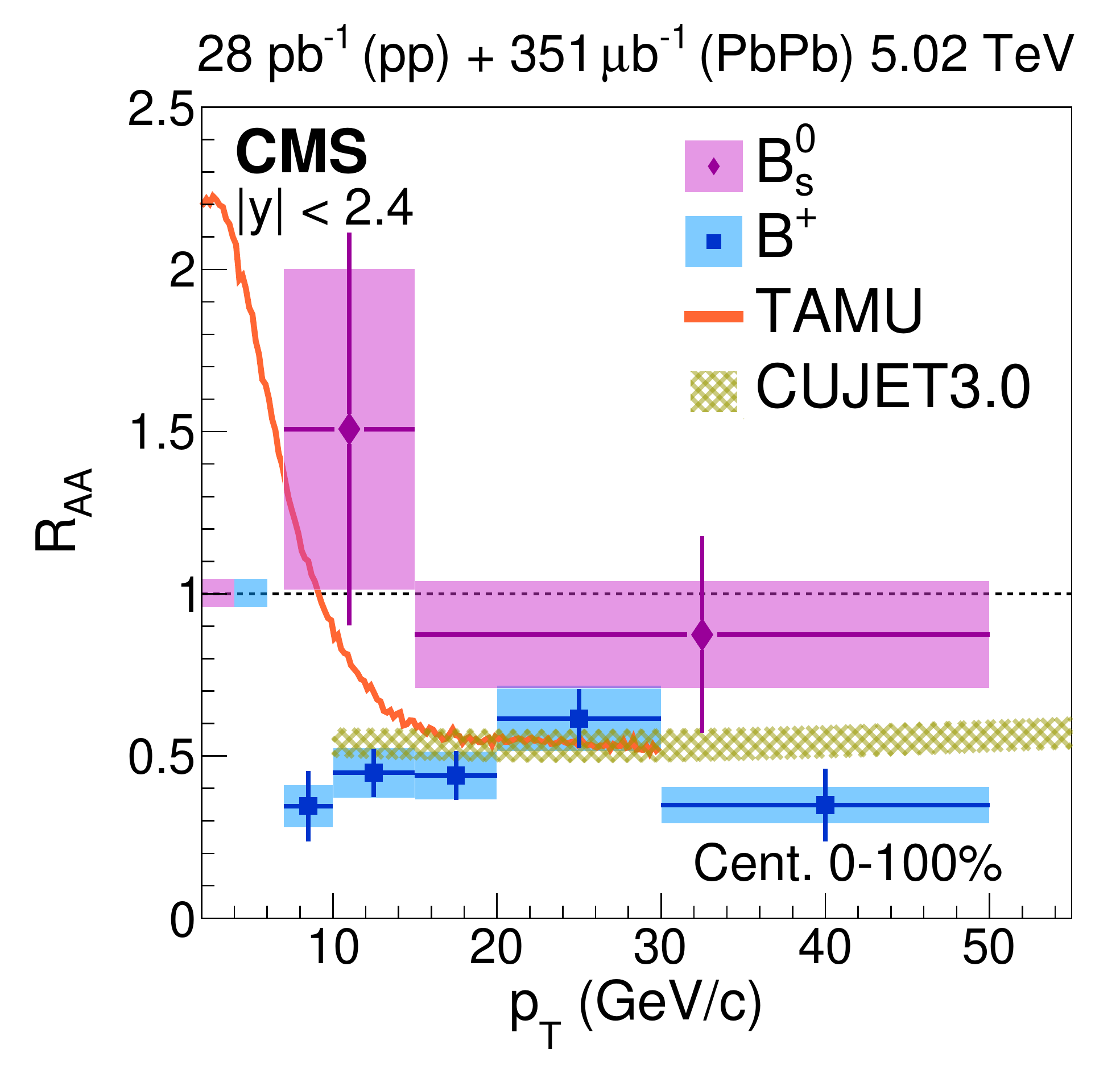}
\caption{Left panel: $B^+$ $R_{AA}$ as a function of $p_T$ \cite{2015BP} and comparison to several theoretical models \cite{CUJET, TAMU, Djordjevic, AdSCFT1, AdSCFT2}. Right panel: $B^0_s$ $R_{AA}$ as a function of $p_{T}$ and comparison with TAMU and CUJET 3.0 \cite{2015Bs}.}
\label{fig:2015BmesonResults}
\end{center}
\end{figure}

The detailed B-meson analysis strategies and procedures using the 2015 pp and PbPb datasets are documented in references\cite{2015BP, 2015Bs}. The final $B^+$ and $B^0_s$ $R_{AA}$ results are shown in Figure \ref{fig:2015BmesonResults}.

Both $B^+$ and $B^0_s$ are suppressed, which suggests that beauty quarks lose energy in the QGP medium. There is no significant $p_T$ dependence of $B^+$ $R_{AA}$ within uncertainties. Our $B^+$ $R_{AA}$ results agree reasonably well with the theoretical predictions. The suppression of $B_s^0$ $R_{AA}$ is less than $B^+$, which may potentially be due to strangeness enhancement. However, the uncertainties are still too high to draw a conclusion.

\section{2018 $B_{s}^{0}/B^{+}$ Analysis Strategies and Procedures}
The same B-meson decay channels as the published 2015 data analysis are used. $B^+$ mesons are fully reconstructed from the decay channel $B^+ \rightarrow J/\psi K^+ \rightarrow \mu^+ \mu^- K^+$ (decay branching fraction = $6.02 \times 10^{-5}$) \cite{PDG2018} and the $B^{0}_{s}$ decay are fully reconstructed from $B^0_s \rightarrow J/\psi \phi \rightarrow \mu^+ \mu^- K^+ K^-$ (decay branching fraction = $3.12 \times 10^{-5}$) \cite{PDG2018}. In B-meson reconstructions, we do not use hadronic particle identifications for kaons. Thanks to CMS precise vertexing and tracking capability, with statistically enriched and dedicated di-muon triggered datasets, clear B-meson signals above the combinatorial background are seen. Moreover, the 2018 dataset has 3 times higher statistics than the 2015 datasets, which allows us to perform more precise and differential measurements.

Then, we apply a Boosted Decision Tree (BDT) algorithm to optimize B-meson signals. We use B-meson Monte Carlo (MC) simulations as signal and a side-band defined from B-meson candidate invariant mass in data as background. We choose the working point based on the maximum statistical significance $\frac{S}{\sqrt{S+B}}$ for each $p_T$ bin. In our BDT training, we use the B-meson decay topological variables and make sure that they are uncorrelated to the reconstructed B-meson mass.

\begin{figure}[hbtp]
\begin{center}
\includegraphics[width=0.30\textwidth]{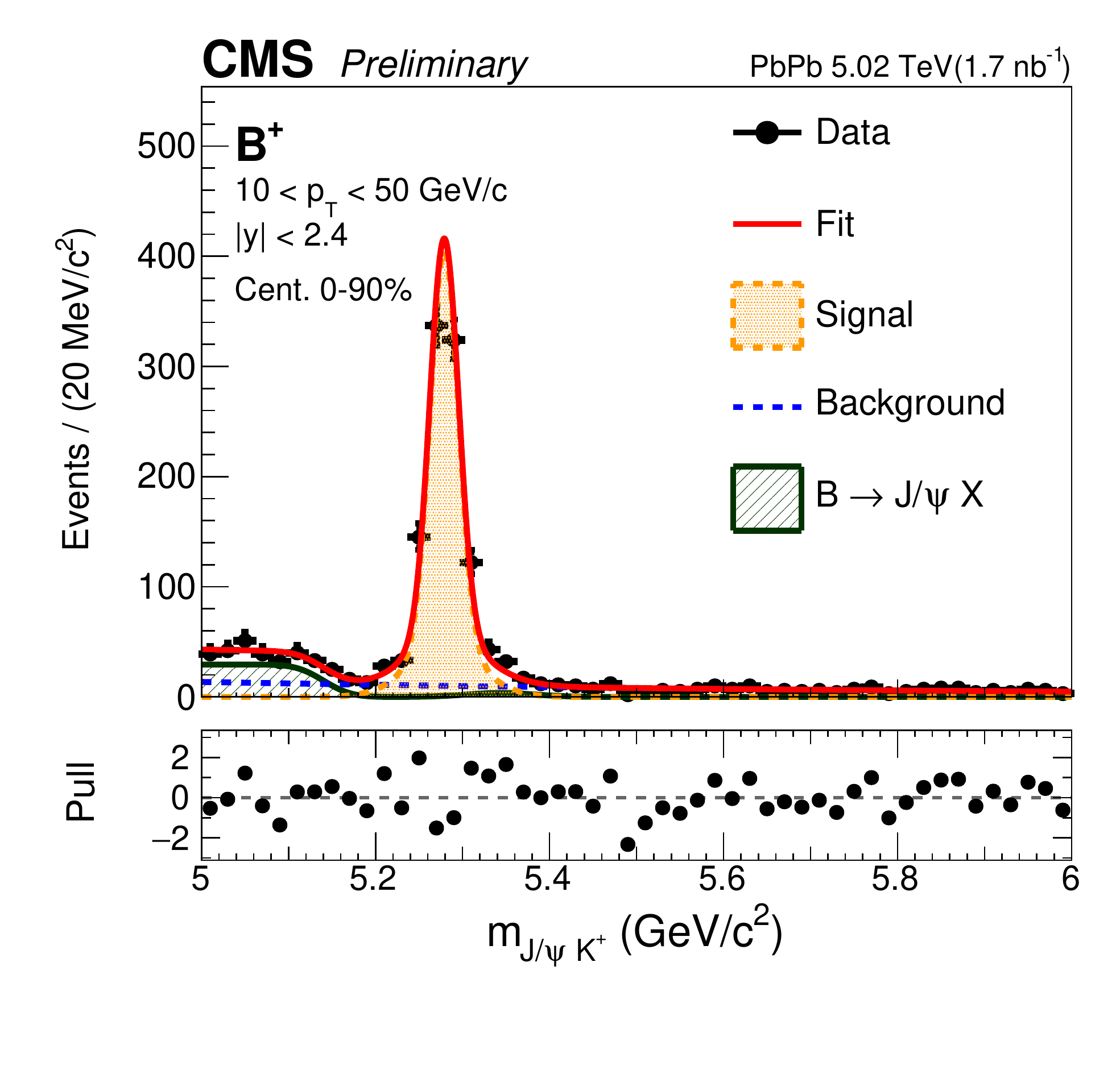}
\includegraphics[width=0.30\textwidth]{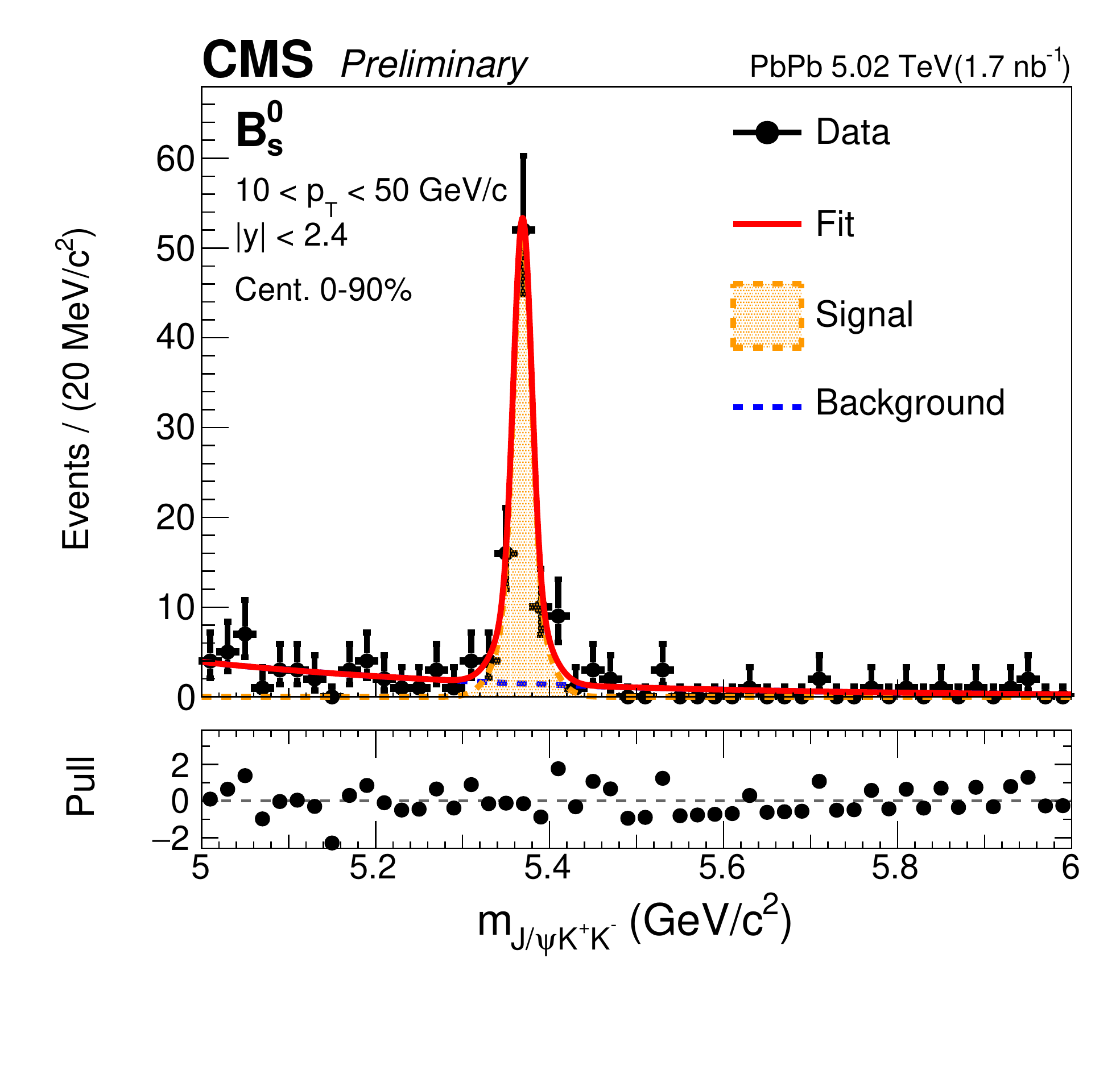}
\caption{The fits on the invariant mass distributions of the fully reconstructed $B^+$ (left) and $B^0_s$ (right) in the $p_T$ range of 10 -- 50 GeV/c and centrality range of 0 -- 90\% in PbPb collisions are shown above.}
\label{fig:BFit1050}
\end{center}
\end{figure}

After applying optimal selections to our data sample, we perform unbinned fits on the B-meson invariant mass to extract the raw yield. We use double Gaussian with the same mean to model the signal, exponential decay for combinatorial background, and error function plus two-sided Gaussian to describe background from decay of other b-hadrons. Our fit results for the inclusive $p_T$ range from 10 -- 50 GeV/c are shown in Figure \ref{fig:BFit1050}. 

Clear $B^+$ and $B^0_s$ are observed. The $B_s^0$ signal has greater than 5$\sigma$ significance. In addition, these results demonstrate that CMS has the capabilities of measuring low $p_T$ B mesons. From the fits, we are able to obtain the B-meson signal raw yield.

Next, we use MC to calculate the efficiency correction factor $\langle \frac{1}{\alpha \times \epsilon} \rangle$. Due to the limited detector acceptance, a fiducial region of $|y| >$ 1.5 is used for $p_{T} < $ 10 GeV/c. To minimize the effect on efficiency correction from B meson $p_T$ spectra shape in a data-driven way, we employ finely grained $\langle \frac{1}{\alpha \times \epsilon} \rangle$ 2D B meson $p_T$ and $|y|$ map. We also apply data-driven method for $J/\psi$ efficiency correction based on the muon selections. We then compute $\langle \frac{1}{\alpha \times \epsilon} \rangle$ using the data B-meson candidates within the signal region according to their kinematic properties in the 2D maps.

\section{2018 $B^0_s/B^+$ Results} 

\begin{figure}[hbtp]
\begin{center}
\includegraphics[width=0.30\textwidth]{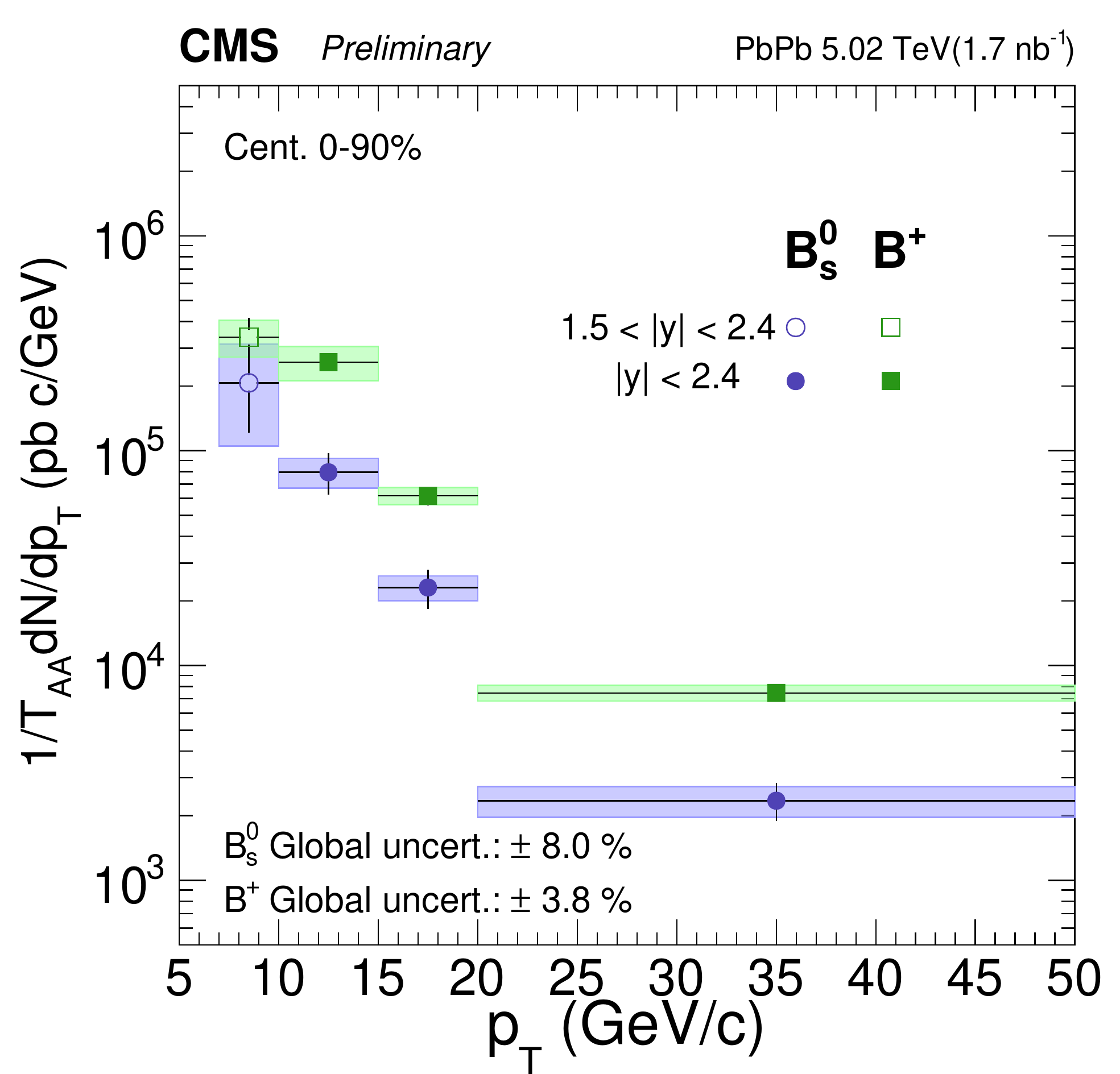}
\includegraphics[width=0.30\textwidth]{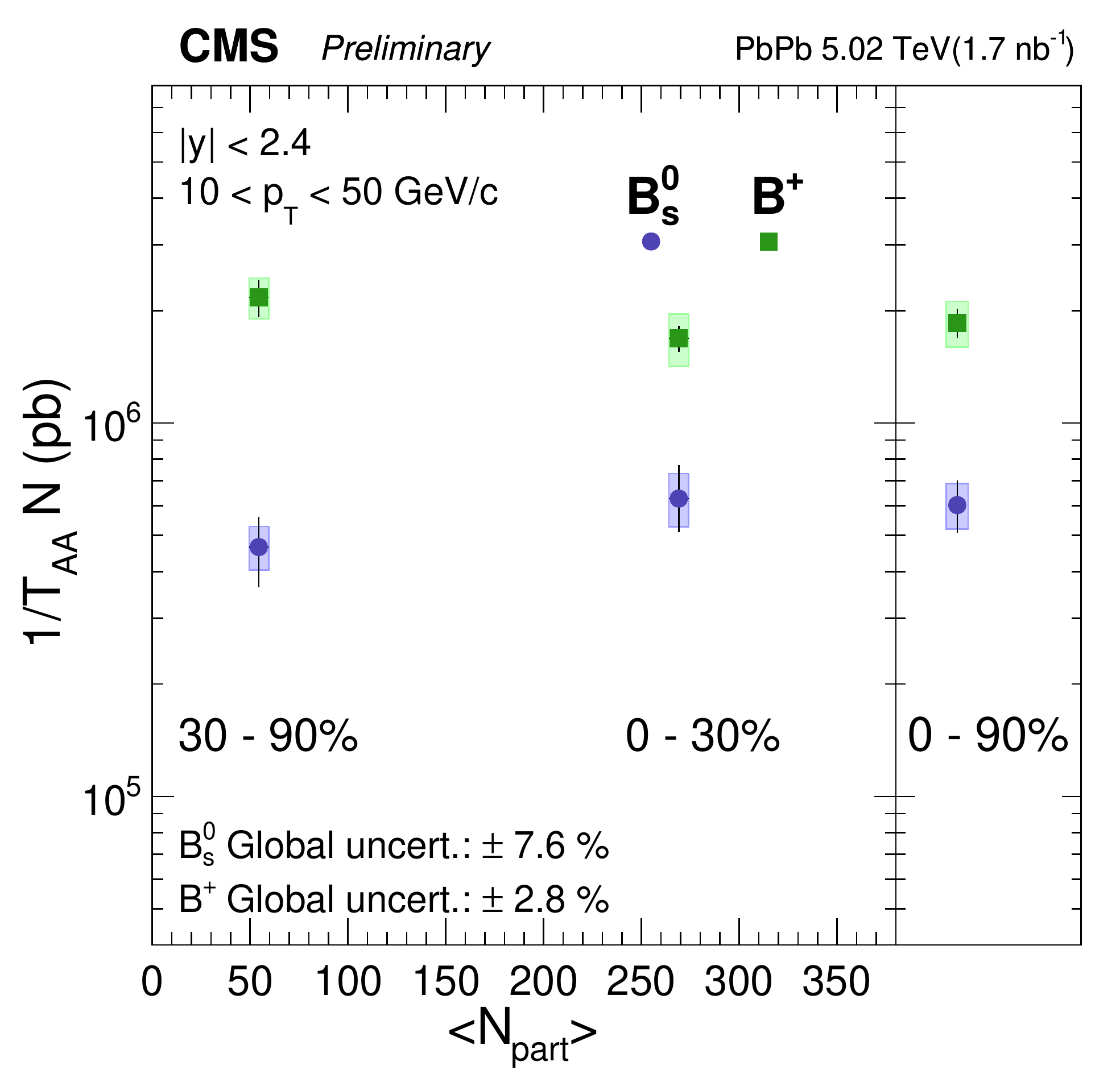}
\caption{The $B^+$ (green) and $B^0_s$ (blue) cross section vs $p_T$ in 0 -- 90\% centrality (left) and vs centrality (right) in 10 -- 50 GeV/c $p_T$ (right) are shown above.}
\label{fig:BXsec}
\end{center}
\end{figure}

The B-meson cross section as a function of transverse momentum and event centrality are shown in Figure \ref{fig:BXsec}.

We have performed precise $p_T$ differential cross-section measurements for $B^+$ and $B^0_s$ mesons from 7 -- 50 GeV/c. This is also our first centrality differential B-meson cross-section measurement. The uncertainties of measurement are high for $B^0_s$ in 7 -- 10 GeV/c due to the limited data sample. 

\begin{figure}[hbtp]
\begin{center}
\includegraphics[width=0.30\textwidth]{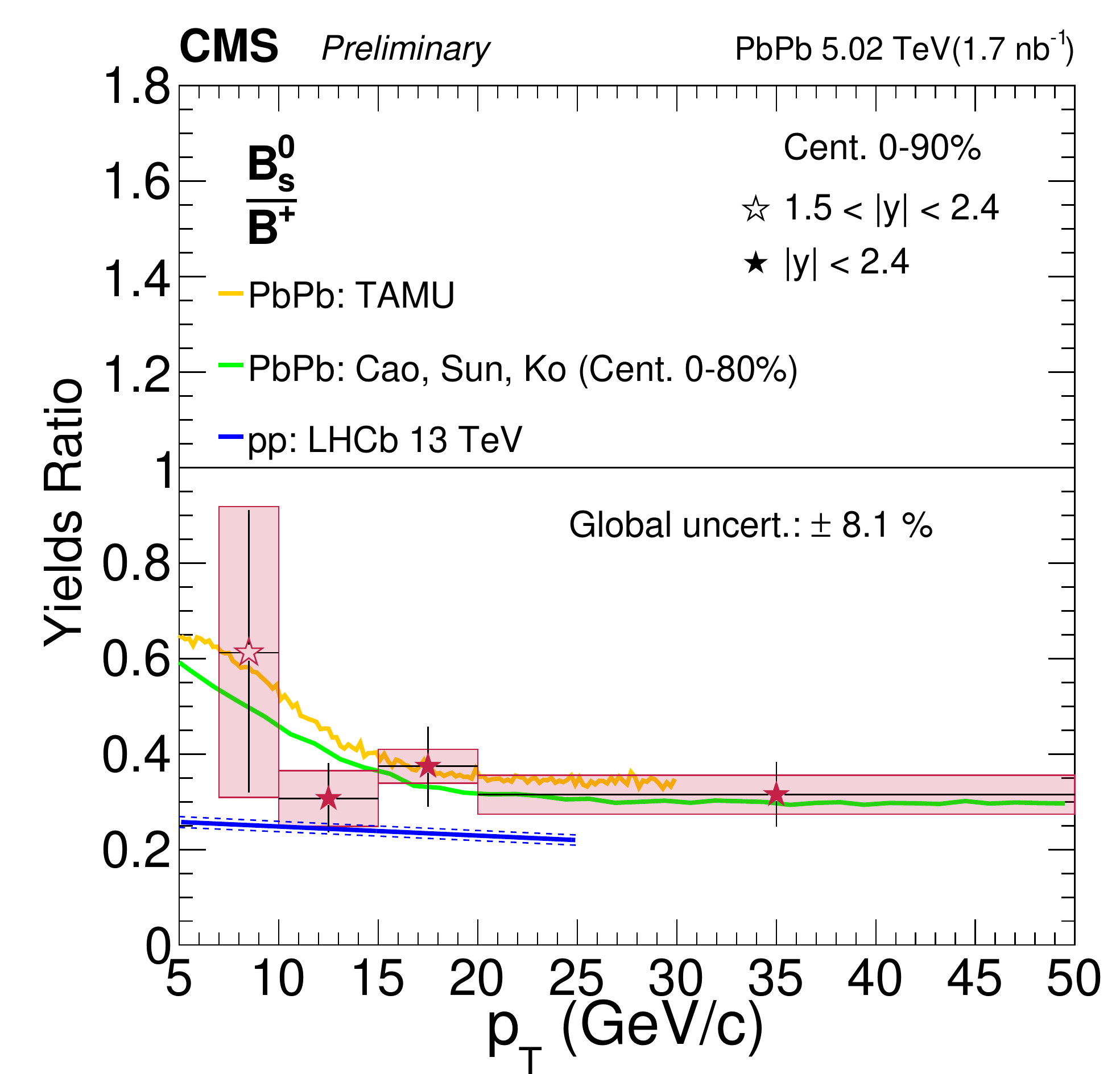}
\includegraphics[width=0.30\textwidth]{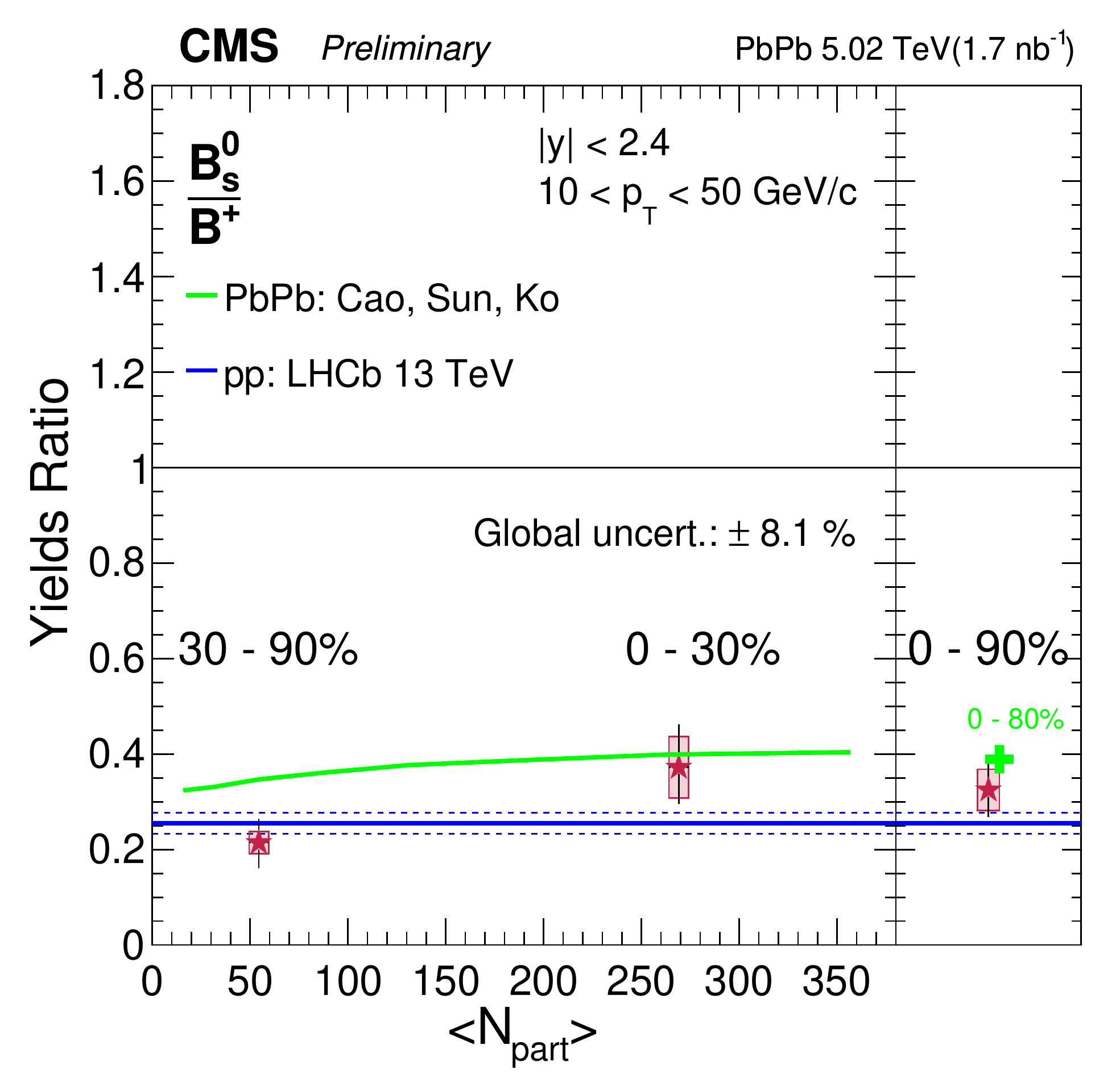}
\caption{The left panel shows the $B^0_s/B^+$ ratio vs $p_T$ in 0 -- 90\% centrality and the comparisons with the LHCb pp reference \cite{LHCbpp}, the TAMU model \cite{TAMU} in PbPb, and the Cao, Sun, Ko model \cite{Cao} in PbPb. The right panel shows the $B^0_s/B^+$ ratio vs centrality in 10 -- 50 GeV/c $p_T$ and the comparisons with LHCb pp reference and the Cao, Sun, Ko model in PbPb.}
\label{fig:BsBPRatio2018}
\end{center}
\end{figure}

Next, we take the ratio of $B^0_s$ cross section to $B^+$ and cancel some systematic uncertainties. Figure \ref{fig:BsBPRatio2018} shows the $B^0_s/B^+$ and the comparison with theoretical predictions and the LHCb pp reference.  

No significant $p_T$ dependence of $B^0_s/B^+$ is observed within current uncertainties. It agrees reasonably well with both the TAMU model (collisional energy loss and heavy quark diffusion in the medium and recombination) \cite{TAMU}  and the Cao, Sun, Ko model (advanced Langevin hydrodynamics model with both heavy quark elastic and inelastic energy loss) \cite{Cao}. The $B^0_s/B^+$ is systematically higher than the LHCb pp 13 TeV data \cite{LHCbpp} within 1$\sigma$. However, it is not significantly enough to conclude strangeness enhancement. 

The $B^0_s/B^+$ vs centrality results are consistent with the Cao, Sun, Ko model except for the peripheral 30 -- 90\% centrality region, which is below the theoretical prediction. It is compatible with the LHCb pp 13 TeV data. There is no strong evidence to support strangeness enhancement for beauty quark hadronization in the QGP medium within our current uncertainties.

\section{Summary}

We have reported the B-meson cross section, $R_{AA}$, and $B^0_s/B^+$ ratio measurements from CMS 2015 and 2018 datasets. Our 2015 $R_{AA}$ results indicate that beauty quarks lose energy in the QGP medium. The first $B^0_s$ signal with a significance greater than 5$\sigma$ in heavy-ion collisions is observed. In addition, we see no significant $p_T$ dependence in the $B^0_s/B^+$ ratio. Both TAMU and Cao, Sun, Ko models agree reasonably well with our $B^0_s/B^+$ data. Our $B^0_s/B^+$ ratio in PbPb is compatible to the pp one measured by LHCb. Given the current uncertainties, the strangeness enhancement of beauty quark hadronization in the QGP medium is still uncertain. Therefore, more precise B-meson measurements will need to be carried out with LHC Run 3 and Run 4 datasets.

\end{document}